\documentclass[prb, twocolumn,showpacs,preprintnumbers,amsmath,amssymb,floatfix,bbm]{revtex4}

\usepackage{graphicx}
\usepackage{dcolumn}
\usepackage{bm}
\usepackage{amsmath,amssymb,bbm}

\begin{document}  
  
\input epsf.sty  
  


\title{Equilibrium entanglement vanishes at finite temperature}  
    
\author{B. V. Fine$^{1,2}$\email{fine@mpipks-dresden.mpg.de}, F. Mintert$^1$, and
A. Buchleitner$^1$}     
   
\affiliation{  
$^1$ Max Planck Institute for the Physics of Complex Systems,  
Noethnitzer Str. 38, D-01187, Dresden, Germany;
\\
$^2$ Physics Department, University of Tennessee,
101 South College,
1413 Circle Drive
Knoxville, TN 37996-1508
USA}  
  
\date{February 1, 2005}

\begin{abstract} 
We show that the equilibrium entanglement of a bipartite
system having a finite number of quantum states 
vanishes at finite temperature, for arbitrary interactions
between its constituents and with the environment.
\end{abstract}  
  
  
\maketitle  
  
  
The interplay between the properties of condensed matter systems  
and the notion of entanglement has recently come into the focus of   
active   
research\cite{Arnesen-etal-01,Wang-01A,Gunlycke-etal-01,Osborne-etal-02,Osterloh-etal-02,Zanardi-etal-02,Kamta-etal-02,Ghosh-etal-03,Vidal-etal-04,Syljuasen-03,Shi-03,Gu-etal-04}.      
In the present paper, we exploit the general topology
of state space and, thereby, formulate a rigorous statement on 
the disappearance of entanglement in equilibrium   
condensed matter systems as a function of temperature.
Below, in order to be specific,
we first consider the 
simplest case of two quantum two-level systems (TLS's),
and then generalize 
the result to the case of bipartite systems with arbitrary [finite] dimension
of the Hilbert space.
  
Consider two TLS's interacting with each other and with   
some environment. The coupling to the environment may be strong or weak.  
Typical examples are: two coupled spins 1/2 belonging to a larger spin lattice,  
or two two-level defects in 
the 
an environment of phonons. The two TLS's   
may become entangled with each other either as a result of direct   
interaction, or via interaction with their common environment.  
We focus on the family of the reduced equilibrium density matrices of   
these two TLS's obtained by tracing out the environmental   
degrees of freedom.   
The entanglement present in the  
equilibrium density matrix of two TLS's will be called ``equilibrium
entanglement''. 
 
In principle, there exist theoretical settings in which  
equilibrium entanglement  
is absent at all  
temperatures.  
However,  
in another very common
situation, when two interacting TLS's are entangled at sufficiently 
low temperatures, the following statement holds:
\begin{quote}
{\em Equilibrium entanglement between two TLS's always vanishes
at a finite temperature.} 
\end{quote} 
The constructive meaning of this 
observation is that the disappearance of   
entanglement never 
exhibits the character of 
continuous crossover   
extending to 
the 
infinite temperature. 
This is consistent with all 
results so far reported on specific physical systems which admit
an analytical or 
numerical computation of some entanglement measure
\cite{Nielsen-98,Arnesen-etal-01,Wang-01,Wang-01A,Kamta-etal-02,Zanardi-etal-02}.   
In general, however, the accurate calculation of the reduced  density  
matrix for two TLS's, which strongly interact with the environment, 
is either very difficult or impossible.   
Notwithstanding, the above statement remains valid, no matter how complex the  
TLS's-plus-environment Hamiltonian   
may be. Therefore, it is applicable   
to all kinds of \mbox{\it experimentally relevant} TLS's.  

It is also worth mentioning that the rule formulated above  
does not constitute a piece of common knowledge  
in the field of condensed matter physics.   
Unlike the case of pure states, the notion of entanglement   
for mixed state density  
matrices is not intuitive, as it discriminates correlations  
which can be described classically from those which cannot.  
A priori, one may have a conflicting intuition about the onset of   
entanglement, since, on the one hand, it is associated with some   
kind of quantum coherence, which may require a phase transition.  
On the other hand, if we focus on the case of two two-level   
defects surrounded  
by a phonon bath,  
the finiteness of the problem excludes the notion of a phase  
transition. 
In turn, this 
suggests that the onset of entanglement   
may have a 
a crossover-like character, extending to the infinite temperature.  

\

Our statement is based on the following argument: 

Let us 
consider a family of $4 \times 4$ equilibrium density
matrices $\varrho$ defined in the basis 
\mbox{
$\{ |\uparrow, \uparrow \rangle, |\uparrow, \downarrow \rangle, 
|\downarrow, \uparrow \rangle, |\downarrow, \downarrow \rangle \}$}, where 
arrows $\uparrow$ and $\downarrow$ encode the 
two states of each TLS. 
As a measure of entanglement, we use concurrence, which can be calculated
as\cite{Wooters-98}
\begin{equation}
c(\varrho) = \hbox{max}[\lambda_1 - \lambda_2 - \lambda_3 - \lambda_4, 0],
\label{C}
\end{equation}
where $\lambda_i$ are the square roots of the eigenvalues of
\begin{equation}
\varrho \sigma_{1y} \sigma_{2y} \varrho^{\ast} \sigma_{1y} \sigma_{2y}, 
\label{rhorho}
\end{equation}
and $\lambda_1$ is the largest among them. Here $\varrho^{\ast}$
is the complex conjugate of $\varrho$, and $\sigma_{1y}$
and $\sigma_{2y}$ are the second Pauli matrices 
$\left( \begin{array}{cc} 0 & -i \\ i & 0  \end{array} \right)$ 
acting on the first and the second TLS, respectively. 
The two TLS's are entangled only when \mbox{$c >0$},
which, in turn, is only possible when 
\mbox{$\lambda_1 - \lambda_2 - \lambda_3 - \lambda_4 > 0$}.

At infinite temperature, the equilibrium density matrix has a particularly 
simple form:
\begin{equation}
\varrho_{\infty} = {\mathbbm 1}/4,
\label{rhoinf}
\end{equation}   
which does not depend on the 
form of the TLS's-plus-environment   
Hamiltonian.  
In practice, the infinite temperature limit is reached when the  
temperature is much higher than the typical energy per one of the   
TLS's. (This energy is the sum of the interaction energies with the other TLS  
and with the environment.)   

Upon substitution of 
$\varrho_{\infty}$ given by Eq.~(\ref{rhoinf}) 
into 
Eq.(\ref{rhorho}), 
we obtain a matrix 
of the form ${\mathbbm 1}/16$,
such that $\lambda_i=1/4$ \mbox{$(i=1,\hdots,4)$},  
and
\mbox{$\lambda_1-\lambda_2-\lambda_3-\lambda_4=-1/2$}.  
Since the values of   
$\lambda_i$ depend continuously on the matrix elements of $\varrho$, 
which, in turn, have a continuous  
dependence on the inverse temperature $1/T$, the value of   
$\lambda_1 - \lambda_2 - \lambda_3 - \lambda_4 $ can 
only turn positive
after increasing the inverse temperature 
from zero to a finite  
value. Given Eq.(\ref{C}), this is the precise equivalent of what
was stated initially. 

\

The generalisation  
to equilibrium entanglement of bipartite quantum systems of arbitrary finite
dimension  
follows from the topological 
observation\cite{Zyczkowski-etal-98} 
stating that, 
in the space of   
all possible $N \times N$ density matrices, the [infinite
temperature] density matrix ${\mathbbm 1}/N$ is surrounded
by a region of separable quantum states, for any bipartite decomposition of the N-level
system. (See also \cite{Braunstein-etal-99}.)
Therefore, an infinite temperature density matrix
cannot be transformed into an entangled density matrix   
by an infinitesimal change of its
matrix elements.

\

A natural question arising in the above context is:  
does the entanglement transition imply any observable effect?   
Promising evidence  
for one such an effect 
has   
recently emerged from the experimental and theoretical 
study \cite{Ghosh-etal-03} of   
insulating magnetic salt LiHo$_x$Y$_{1-x}$F$_4$, which, for our present  
theoretical purposes, can be described as a set  
of spins 1/2 randomly placed on a crystal lattice and coupled to   
each other by magnetic dipolar interaction.  
This study has shown that the onset of entanglement may explain   
the qualitative difference between the structured behavior of specific heat  
and the featureless behavior of magnetic susceptibility. It was also shown
that the onset of entanglement is accompanied by rapidly growing deviations  
between 
predictions from classical and quantum theory.  

Another interesting possibility is that the entanglement transition
may signify a limit beyond which various high-temperature expansion
techniques become unreliable. Since any high-temperature expansion starts from
the infinite temperature, and since there is no entanglement in a finite
range around infinite temperature, it remains an open question whether
such techniques can possibly generate an entangled
density matrix at finite temperature.

In conclusion, we 
hope 
that the fact that, in a large class 
of condensed matter systems, the entanglement transition occurs  
at finite temperature    
will stimulate further 
experimental and theoretical efforts to characterize this transition.

\bibliography{entgl}   
  
  
  
  

\end{document}